\documentclass{optica-article}

\journal{opticajournal} 

\articletype{Research Article}

\usepackage{lineno}

\usepackage{graphicx}
\usepackage{dcolumn}
\usepackage{bm}
\usepackage{siunitx}

\usepackage{physics}
\usepackage{mathrsfs}
\usepackage{multirow}

\newcommand{\QOT}{Centre for Quantum Optical Technologies, Centre of New Technologies, University of Warsaw, Banacha 2c, 02-097 Warsaw, Poland}

\newcommand{\FUW}{Faculty of Physics, University of Warsaw, Pasteura 5, 02-093 Warsaw, Poland}

\newcommand{\ulfasr}{SUSI}

\begin{document}

\title{Super-resolution of ultrafast pulses via spectral inversion}

\author{Michał Lipka,\authormark{1,2,*} and Michał Parniak,\authormark{1,2}}

\address{\authormark{1}\QOT}
\address{\authormark{2}\FUW}
\email{\authormark{*}mj.lipka@uw.edu.pl}

\begin{abstract*}
    The resolution limits of classical spectroscopy can be surpassed by quantum-inspired methods leveraging the information contained in the phase of the complex electromagnetic field. Their counterpart in spatial imaging has been widely discussed and demonstrated; however, the spectral-domain implementations are few and scarce.
    We experimentally demonstrate a spectroscopic super-resolution method aimed at broadband light (10s to 100s of GHz), and based on the spectral-domain analog of image inversion interferometry. In a proof-of-principle experiment, we study the paradigmatic problem of estimating a small separation between two incoherent spectral features of equal brightness, with a small number of photons per coherence time. On the grounds of asymptotic estimation theory, more than a $2$-fold improvement over the spectral direct imaging is demonstrated in terms of required resources (photons) for a given estimator variance. The setup is based on an actively stabilized Mach-Zehnder-type interferometer with electro-optic time lenses and passive spectral dispersers implementing the inversion. As such, the method promises on-chip integration, good scalability, and further applications e.g. for mode sorting. 
\end{abstract*}

\section{Introduction}

The concept of resolution inseparably accompanies any measurement technique. 
While very intuitive, its rigorous description long remained surprisingly elusive with a practical formulation dating back to Lord Rayleigh \cite{Rayleigh1879} and known as the Rayleigh limit, together with its spectroscopic counterpart \textendash{} the Fourier limit \cite{moller-1971}. The descriptions, similar in essence, referred to a minimal separation between two identical features at which they could still be distinguished.
While a modern analysis on the grounds of asymptotic parameter estimation theory brought quantitative answers \cite{bettens1999modelbased}, a truly groundbreaking finding was how to overcome these limits.

Traditional direct imaging or spectroscopy employs quadratic-response photodetectors which output is proportional to the image-plane intensity $\propto \abs{E(f)}^2$. In a seminal paper Tsang \textit{et al.} \cite{Tsang2016} demonstrated the potential for using the information encoded in the phase of the complex electric field $E(f)$, inherently lost in direct imaging, to overcome the Rayleigh limit. This finding inspired numerous theoretical proposals and experimental super-resolution demonstrations \cite{Paur2016, Yang2016, Nair2016, Tang2016fault, Tham2017beating, Parniak2018, larson2019common, Datta2020,  Frank2023, Jordan2023, Triggiani2023, Paur2018tempering,  Backlund2018, rehacek2019intensity, Boucher2020spatial, syamsundar2021effects, Zanforlin2022optical, frank2023passive, wadood2024experimental, sajjad2021attaining, sorelli2021optimal, wang2023fundamental, Mitchell2023}.

From a theoretical point of view, space-time analogy \cite{Torres2011} makes the problem of spectral resolution almost equivalent to 1-dimensional spatial imaging \cite{Rudnicki2024spectral}. However, surprisingly, there are few experimental implementations of super-resolution spectral imaging. These include a quantum-pulse-gate-based super-resolution in the spectral \cite{Donohue2018} and temporal \cite{Ansari2021} domains,
heterodyne detection with post-processing \cite{krokosz2023beating}, and a time-axis-inversion-based protocol implemented in a quantum memory \cite{Mazelanik2022}. The latter is based on the same principle as this work, namely of image inversion interferometry, however, the implementation inherently targets tens of \SI{}{\kilo\hertz} bandwidth pulses and hence differs in almost every aspect. 

While we focus on weak and incoherent light, such as encountered in fluorescence spectroscopy \textendash{} a powerful and vastly employed tool of organic chemistry, biology, and medicine \cite{bose2018fluorescence}, different spectroscopic settings merit individual approach. For instance super-resolution has been demonstrated for laser spectroscopy \cite{Boschetti2020} and with controlled illumination spectroscopic methods can be vastly enhanced with quantum light probes \cite{Mukamel2020, Lipka2021prl,  hashimoto2024fourier, grenapin2023superresolution}.

In this work, we present a spectral-domain super-resolution method inspired by the image inversion interferometry \textendash{} a technique well described and demonstrated in the context of super-resolution in spatial imaging \cite{Wicker2009character, Nair2016, Tang2016fault, larson2019common}.
We call the method SUper-resolution via Spectral Inversion (\ulfasr{}). In an analogy to its spatial counterpart, \ulfasr{} is based on interfering an inverted copy of the input state with its original version. The inversion operator $\mathscr{P}$ acts in the spectral degree of freedom, around an \textit{a priori} known centroid. The symmetric ($+$) and antisymmetric ($-$) parts of the state are eigenvectors of $\mathscr{P}$ with $\pm1$ eigenvalues, hence in the interferometric setup the ($+$) and ($-$) parts can be spatially separated and independently undergo photon counting. If the input state consists of an incoherent mixture of closely separated equally bright pulses, almost all information on their separation is contained in the number of photons counted in the ($-$) port and there is no inherent background i.e. for 0 separation, with an ideal setup, no photons are going to the ($-$) port. In this way \ulfasr{} provides a sensitive and hence efficient estimation in the sub-Fourier regime.  

The super-resolution methods essentially perform mode sorting in a scenario-specific basis of modes. Often in practice, the basis is truncated to the first two modes which are nevertheless sufficient in the regime of very small separations.  
Promising multi-mode-sorting techniques have been described \cite{horoshko2023interferometric} and demonstrated \cite{serino2023realization} in the spectral domain, with applications beyond parameter estimation. \ulfasr{} could be in principle extended to multi-mode sorting with a network of spectral-inversion interferometers and linear optics \textendash{} a promising scheme if combined with on-chip integration.

The article is organized as follows: in sec.~\ref{sec:theory} we briefly introduce the problem and \ulfasr{} in mathematical terms and consider a simplified case of Gaussian pulses; in sec.~\ref{sec:results} experimental results are presented; sec.~\ref{sec:methods} describes the methodology, and finally sec.~\ref{sec:discussion} and sec.~\ref{sec:conclusions} give a conclusion and outlook. 
\begin{figure}[ht!]
\centering\includegraphics[width=0.8\columnwidth]{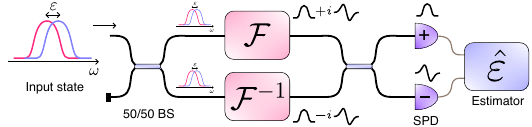}
\caption{
The idea of \ulfasr{}. An incoherent mixture of spectrally separated (by $\varepsilon$) pulses, with \textit{a priori} known centroid, enters a Mach-Zehnder-type interferometer. The combined action of a Fourier transform $\mathscr{F}$ in one of its arms and an inverse Fourier transform $\mathscr{F}^{-1}$ in the other is equivalent to a spectral inversion $\mathscr{P}$ in a single arm. Interfering the signals from both arms separates the symmetric ($+$) and antisymmetric ($-$) part of the input mode. Photon counting at the output ports performs a more sensitive sensing of $\varepsilon$ (lower estimator variance) than direct imaging (spectroscopy), hence constituting a super-resolution method.
50/50 BS - balanced beam splitter, SPD - single photon detector.
}
\label{fig:idea}
\end{figure}
\section{Theory\label{sec:theory}}
In a standard spectroscopic approach estimating the separation between two spectral features would involve detecting a statistic of $N$ photons with a spectrally-resolving detector. The number of photons in each spectral bin would be used to construct e.g. a maximum likelihood estimator of the separation under an assumed statistical model.

Increasing $N$ will lower the estimator variance. Hence, even for spectral features of width $\sigma$ separated by $\delta f$ such that $\varepsilon=\delta f/\sigma \ll 1$ a standard approach (direct imaging) is still possible, if not practically feasible.

Inevitably the problem must be formulated in terms of required resources. Here a meaningful choice is the number of observed photons $N$, which usually corresponds to the measurement time. A super-resolution method can lower the number of resources required for a given estimator variance, with all other factors constant. In comparative terms, super-resolution methods excel for $\varepsilon\ll 1$ and henceforth we shall assume this sub-Fourier regime.

In the paradigm of frequentist inference, a modern and convenient evaluation of an estimation scheme comes from the Cramér–Rao bound and Fisher information \cite{Kay1993}.
The variance of any unbiased estimator $\hat{\varepsilon}$ follows the Cramér–Rao bound: 
\begin{equation}
    \Delta^2\hat{\varepsilon}\geq \frac{1}{F(\varepsilon)},
\end{equation}
where $F(\varepsilon)$ is the Fisher information of estimating a parameter $\varepsilon$ in a measurement scheme given by a set of outcomes $\lbrace i \rbrace$ associated with conditional probabilities $\lbrace P(i|\varepsilon)\rbrace$. $F(\varepsilon)$ is given by
\begin{equation}
\label{eq:fisher}
    F(\varepsilon) = \sum_i \frac
    { \left(\partial_\varepsilon P(i|\varepsilon)\right)^2}
    {  P(i|\varepsilon) }.
\end{equation}

Let us note that the problem of estimating the frequency separation of two incoherent pulses is in many ways analogous to the well-studied case of the spatial far-field imaging of two point-like sources. The main difference is as follows. In our case, the width of the pulse is its intrinsic property. It is also the scale parameter of the problem. However, the spectral resolution of any instrument is assumed to be arbitrarily good. On the other hand, for spatial imaging, the width of the imaging instrument's point spread function determines this scale. Of course, considering space-time duality, one can imagine a temporal aperture that gives the spectral shape to the pulses. 
Note that minute changes in the spectral shape of the pulses can drastically broaden its temporal extent leaving the spectrum width $\sigma$ constant. A limiting case is a perfectly square spectrum which would require an infinite temporal width. 
Hence, under such an unconstrained formulation, the DI methods will seemingly have a very good performance for super-Gaussian pulses. Introducing a limited temporal aperture (or equivalently limited spectrometer resolution) would restore a realistic DI behavior. 

We shall follow a standard methodology for the spatial super-resolution methods. Let us assume that the number of photons per coherence time is very low. It is a well-grounded assumption, well discussed in ref.~\cite{Tsang2016} and true for fluorescence spectroscopy and most natural phenomena such as stellar emission \cite{huang2023ultimate}.
This way, the probabilities $\lbrace P(i|\varepsilon)\rbrace$ will be Poisson-distributed with means $\mu_i(\varepsilon)=N p(i|\varepsilon)$, where $p(i|\varepsilon)$ is the equivalent probability for a single photon. Poisson statistic produces the shot-noise scaling of the estimator variance $\Delta^2\hat{\varepsilon}\propto1/N$ or $F\propto N$, hence we will further consider the Fisher information per photon $\mathcal{F}=F/N$, which is given by Eq.~(\ref{eq:fisher}) upon substituting $P(i|\varepsilon)\rightarrow p(i|\varepsilon)$. 

Furthermore, we shall assume fully incoherent and equally bright pulses. Deviations from this simplest scenario are well studied in the spatial imaging setting \cite{Tsang2019Resolving, amato2023single, Santamaria:23, Greenwood_2023, Hu2023}.
We will also assume an \textit{a priori} known centroid of the pulses joint spectrum. In practice, a part of the available photons (or measurement time) would be sacrificed to estimate the centroid, possibly in an adaptive strategy \cite{grace2020approaching}.

For notational convenience, we will use dimensionless quantities with the frequency $\xi=f/\sigma$, and pulse separation $\varepsilon=\delta f/\sigma$. For a single pulse a normalized slowly varying envelope of the electric field will be denoted by $\mathcal{A}(\xi)$ obeying symmetry $\mathcal{A}(\xi)=\mathcal{A}(-\xi)$ and normalization
\begin{equation}
    \int \dd{\xi} \abs{\mathcal{A}(\xi)}^2 \xi^2 = \int \dd{\xi} \abs{\mathcal{A}(\xi)}^2 = 1.
\end{equation}

A convenient feature of the incoherent mixture of two weak pulses $\mathcal{A}(\xi\pm\varepsilon/2)$ is that each pulse can be considered separately. In the end, the photon counts would be added together for $\pm\varepsilon/2$ pulses.

\subsection{Spectral inversion interferometry}
A simplified setup for \ulfasr{} has been depicted in Fig.~\ref{fig:idea}. 
The main part consists of a Mach-Zehnder-type interferometer with a Fourier transform $\mathscr{F}$ in one arm and an inverse Fourier transform $\mathscr{F}^{-1}$ in the other. Since all elements are linear and the Fourier transform does not change the total energy (Parseval's theorem) such a setup is equivalent to placing two Fourier transforms in series in a single arm, and leaving the other arm empty. The composition of two Fourier transforms has the property of inversion $\mathscr{F}\circ\mathscr{F}\left[g(x)\right]=g(-x)$ which implements the spectral equivalent of image inversion interferometry. 
Distributing the transform into two arms has practical merits in ensuring similar efficiency and higher-order imperfections in each arm.

Let us calculate the result of a \ulfasr{} measurement for a single shifted pulse $\mathcal{A}(\xi-\varepsilon/2)$, in the regime of $\varepsilon\ll 1$.
From the pulse expansion
 \begin{equation}
     \mathcal{A}(\xi-\frac{\varepsilon}{2}) = \mathcal{A}(\xi)-\frac{\varepsilon}{2}\partial_\xi \mathcal{A}(\xi)
     +\frac{\varepsilon^2}{8}\partial^2_\xi \mathcal{A}(\xi)
     +\order{\xi^3}.
 \end{equation}
we can see that the information about $\varepsilon$ is contained in the antisymmetric part $\partial_\xi \mathcal{A}(\xi)$.
Neglecting higher-order terms and acting with the (inverse) Fourier transform ($\xi\xrightarrow{}\tau$) we get 
\begin{align}
    \mathscr{F}\left[\mathcal{A}(\xi-\frac{\varepsilon}{2})\right] &= 
    \tilde{\mathcal{A}}(\tau)\left[
    1+i\tau\frac{\varepsilon}{2}
    -\frac{\varepsilon^2\tau^2}{8}
    \right],\label{eq:fourier_shifted}\\
    \mathscr{F}^{-1}\left[\mathcal{A}(\xi-\frac{\varepsilon}{2})\right] &=
    \tilde{\mathcal{A}}(\tau)\left[
    1-i\tau\frac{\varepsilon}{2}
    -\frac{\varepsilon^2\tau^2}{8}
    \right],\label{eq:fourier_inv_shifted}
\end{align}
where $\tilde{\mathcal{A}}=\mathscr{F}\left[\mathcal{A}\right]$. 
Including the effect of non-perfect visibility $0\leq\mathcal{V}<1$ and after the final beamsplitter transformation, we get the following spectral intensities:
\begin{multline}
    I_\pm = \frac{1}{4} \left(
    \abs{
        \mathscr{F}\left[\mathcal{A}(\xi-\frac{\varepsilon}{2})\right]
    }^2
    + \abs{
        \mathscr{F}^{-1}\left[\mathcal{A}(\xi-\frac{\varepsilon}{2})\right]
    }^2 \right) \\
    \pm  \frac{\mathcal{V}}{2} \Re{
        \mathscr{F}\left[\mathcal{A}(\xi-\frac{\varepsilon}{2})\right]    
        \mathscr{F}^{-1}\left[\mathcal{A}(\xi-\frac{\varepsilon}{2})\right]^*
    }.
\end{multline}
Substituting Eqs.~(\ref{eq:fourier_shifted}) and (\ref{eq:fourier_inv_shifted}) gives
\begin{equation}
    I_\pm(\tau) = \abs{\tilde{\mathcal{A}}(\tau)}^2 \left[
    \frac{1\pm\mathcal{V}}{2}
    \mp \frac{\mathcal{V}}{4}\tau^2\varepsilon^2 
    \right]
    +\order{\varepsilon^3}.
\end{equation}
Assuming bucket detectors, we integrate over $\tau$ and drop higher-order $\varepsilon$ terms:
\begin{align}
\label{eq:npm}
    n_\pm &= \int \dd{\tau} I_\pm(\tau) = \frac{1\pm\mathcal{V}}{2} 
    \mp \frac{\mathcal{V}}{4}\varepsilon^2 \mathit{\Delta}
    ,\\
    \mathit{\Delta} &= 
    \int \dd{\tau} \abs{\tilde{\mathcal{A}}(\tau)}^2\tau^2,
\end{align}
where we used Parseval's theorem and $\mathcal{A}(\xi)$ normalization. The variance $\mathit{\Delta}$ depends on the spectrum shape. 
For instance for Gaussian pulses in our dimensionless units, it reads $\mathit{\Delta}_\mathrm{Gauss}=1/4$.

Summing $n_\pm$ for $\pm\varepsilon/2$ pulses and using Eq.~(\ref{eq:fisher}) with $P(i|\varepsilon)\rightarrow p(i|\varepsilon)$ and 
\begin{equation}
\label{eq:ppm}
p(\pm|\varepsilon)=n_\pm/(n_+ + n_-)
\end{equation}
we get Fisher information per photon $\mathcal{F}(\varepsilon) = F(\varepsilon)/N$ of
\begin{equation}
\label{eq:fofeps}
    \mathcal{F}(\varepsilon) = 
    \frac{\mathcal{V}^2}{1-\mathcal{V}^2}
    \frac{\varepsilon^2\mathit{\Delta}^2 }
    {\left( 
    1-\frac{\varepsilon^2\mathit{\Delta}}
    {2} 
    \right)^2}
    = \frac{\mathcal{V}^2\mathit{\Delta}^2}{1-\mathcal{V}^2}\varepsilon^2  +\order{\varepsilon^3}.
\end{equation}

For identical incoherent pulses in an ideal scenario of perfect visibility $\mathcal{V}=1$, the quantum Fisher information (which ultimately bounds the Fisher information for any measurement scheme) remains constant regardless of $\varepsilon$. However, it has been shown that any imperfections restore the $\propto \varepsilon^2$ scaling \cite{oh2021quantum, Schlichtholz2024practical}.
Hence, one meaningful way to compare super-resolution protocols is via the ratio of their quadratic scaling factors. 
Following ref.~\cite{Mazelanik2022}, the super-resolution parameter is defined as
\begin{equation}
    \mathbf{s} = \lim_{\varepsilon\rightarrow{}0}
    \frac{\mathcal{F}_\mathrm{\ulfasr{}}(\varepsilon)}
    {\mathcal{F_\mathrm{DI}}(\varepsilon)}
    \label{eq:superres_param}
\end{equation}
and quantifies the reduction in the number of photons required for a given estimation precision, compared to direct imaging.

For example an analytically tractable case of Gaussian pulses gives $\mathcal{F}_\mathrm{\ulfasr{}}^{(\mathrm{Gauss})}=\varepsilon^2/16\times\mathcal{V}^2/(1-\mathcal{V}^2)$ and $\mathcal{F}_\mathrm{DI}^{(\mathrm{Gauss})}=\varepsilon^2/8$, hence $\mathbf{s}^{(\mathrm{Gauss})}=1/2\times\mathcal{V}^2/(1-\mathcal{V}^2)$. In this case, to perform better than DI, \ulfasr{} would require a visibility of more than ca. $81.7\%$.

Notably, in the limit of $\varepsilon\rightarrow0$ a fraction of $(1+\mathcal{V})/2$ of total $\mathcal{F}_\mathrm{\ulfasr{}}$ comes from the dark ($-$) port counts. 
For instance with our experimental visibility of ca. $\mathcal{V}=93\%$ the dark port provides $96.5\%$ of information.

\section{Results\label{sec:results}}
\begin{figure*}[ht!]
\centering\includegraphics[width=1\columnwidth]{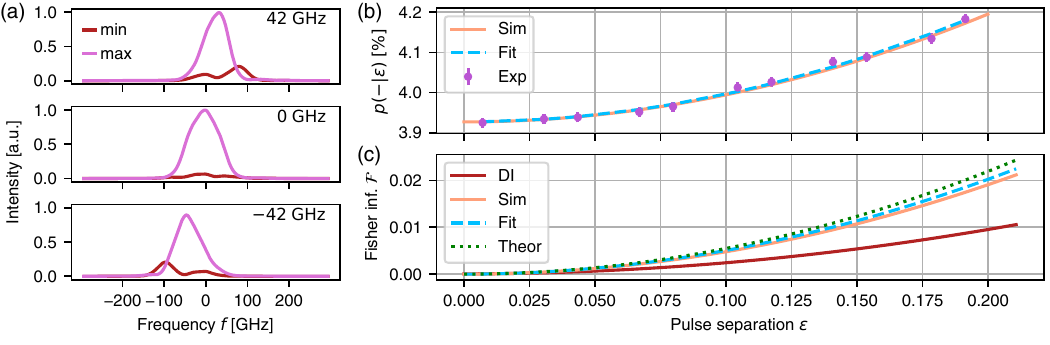}
\caption{
(a) Spectra at the interferometer output for the constructive (max) or destructive (min) interference, with an input pulse frequency-shifted by 0 or \SI{\pm42}{\giga\hertz}.
(b) Fraction of integrated photo counts in the ($-$) port $p(-|\varepsilon)$ during the measurement period, for an incoherent mix of pulses spectrally separated by $\varepsilon$. Experimental data (Exp) with a parabolic fit (Fit) and a result simulated with the measured spectrum and visibility (Sim). 
(c) Fisher information per photon $\mathscr{F}$ for $\varepsilon$ estimation. Fit to experimental data (Fit), analytical prediction (Theor), simulation (Sim), and simulated direct imaging (DI) using the measured spectrum. 
}
\label{fig:results}
\end{figure*}

Experimental results are depicted in Fig.~\ref{fig:results}. The effect of shifting a pulse from the central frequency of the Fourier transformers can be observed in the spectra, as demonstrated in Fig.~\ref{fig:results} (a). 
In the main measurement, for each $\varepsilon$, temporal histograms of photon counts at $\pm$ ports $h_\pm(t,\varepsilon)$ are collected. Counts are then integrated over the measurement period $n_\pm(\varepsilon)=\int \dd{t}\:h_\pm(t,\varepsilon)$ and summed over experiment repetitions.
Finally, the conditional probability $p(-|\varepsilon)$ is estimated as the fraction of photon counts in the dark port.
A quadratic model of Eqs.~(\ref{eq:npm}) and (\ref{eq:ppm}) is fitted to $p(-|\varepsilon)$ and the corresponding Fisher information is calculated from Eq.~(\ref{eq:fofeps}).
Fig.~\ref{fig:results} (b) depicts experimental $p(-|\varepsilon)$ alongside a quadratic fit and a result from a \ulfasr{} simulation.
Fisher information $\mathcal{F}(\varepsilon)$ for the quadratic fit, \ulfasr{} simulation, simulated direct imaging, and a theoretical prediction of Eq.~(\ref{eq:fofeps}) has been depicted in Fig.~\ref{fig:results} (c). Both \ulfasr{} and DI simulations assume an experimentally measured pulse spectrum and in the case of \ulfasr{} a visibility of $93\%$ (c.f. appendix sec.~\ref{sec:app_simulation}, and \ref{sec:app_di_sim}). For the theoretical prediction $\mathit{\Delta}\approx 0.293$ is calculated from the Fourier transform of the experimental spectrum.

The quadratic fit to experimental data and the simulated DI give a super-resolution parameter of $\mathbf{s}=2.13$ 
The results and parameters have been summarized in Tab.~\ref{tab:params}.

We observe a good agreement of the experimental data with both the theoretical prediction and the numerical simulation. By constructing a measurement model, super-resolution with respect to the spectral DI is demonstrated. The super-resolution is quantified in terms of a higher Fisher information per photon in the regime of low separations $\varepsilon\ll 1$, regardless of $\varepsilon$. Via the Cramér-Rao bound it can be understood as the equivalent reduction in the required number of photons to reach a fixed estimator variance.   

\begin{table}[b]
\small
\centering
\caption{\label{tab:params}%
Summary of results assuming $\mathcal{V}=93\%$. The variance of an individual pulse spectrum $\mathit{\Delta}$ strongly affects the lowest non-vanishing expansion coefficient of the Fisher information $\partial_\varepsilon^2\mathcal{F}(0)/2$. The latter is used to obtain the super-resolution parameter $\mathbf{s}$.
}
\begin{tabular}{rccc}
\multicolumn{1}{l}{} & $\mathit{\Delta}$              & $\partial^2_\varepsilon \mathcal{F}(0)/2$       & $\mathbf{s}$                                            \\ \hline
DI                    & \multirow{4}{*}{$0.293$}       & $0.24$                                                   & 1                                                       \\
Theor                 &                                & $0.55$                                                   & $2.29$                                                  \\
Sim                   &                                & $0.48$                                                   & $2.01$                                                  \\
Fit                   &                                & $0.51$                                                   & $2.13$                                                  \\ \hline
DI Gauss              & \multirow{2}{*}{$1/4$} & $1/8$                                            & 1                                                       \\
Theor Gauss           &                                & $0.40$ & $1.60$
\end{tabular}%
\end{table}
%
%
\section{Methods\label{sec:methods}}

\begin{figure*}[ht!]
\centering\includegraphics[width=1\textwidth]{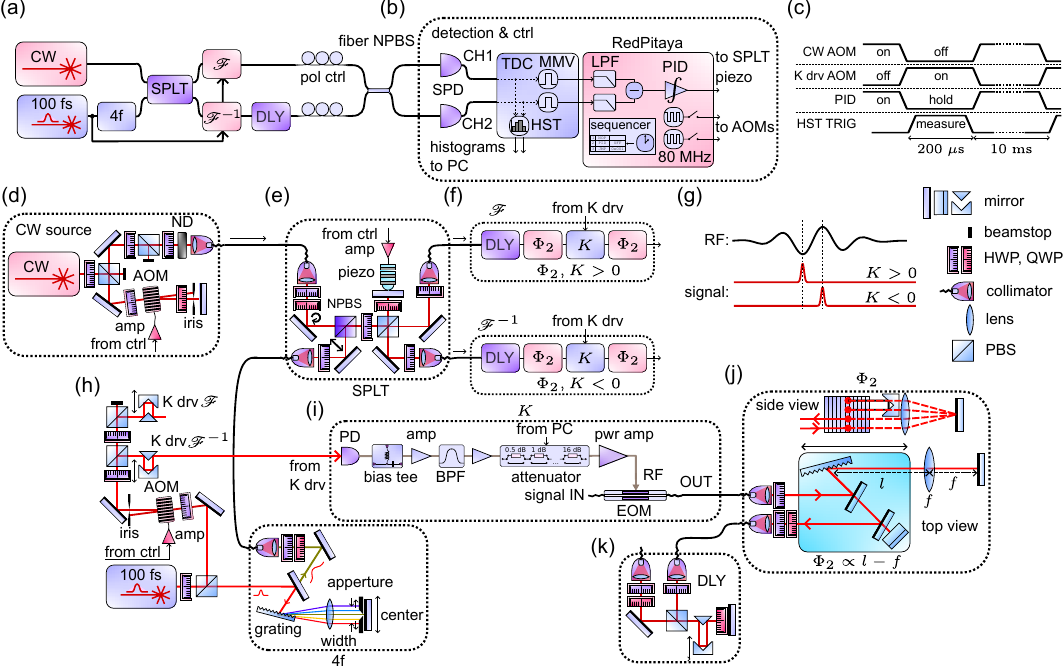}
\caption{
(a)-(j) Experimental setup and experiment sequence.  (a) General schematic. Setup comprises a Mach-Zehnder-type phase-stabilized interferometer with electro-optic (inverse) Fourier transformers in each arm.
(b) Single-photon detection, time-tagging, and histogramming followed by FPGA (RedPitaya) processing. FPGA-based feedback loop stabilizes the difference of photon counts (interferometer phase) 
(c) Experimental sequence controlled via an FPGA sequencer.
(d) CW source with a double-pass acousto-optical modulator (AOM) switch.
(e) Splitting of the signal pulses and combining the signal with a continuous wave (CW) laser used for the feedback loop. 
(f) General schematic of (an inverse) a Fourier transformer $\mathscr{F}$ ($\mathscr{F}^{-1}$). 
(g) Synchronization of a radio frequency (RF) driving signal with an optical pulse, for different signs of the time lens chirp rate $K$.
(i) Electro-optic time lens. The signal from an optically excited photodiode is filtered and amplified to drive an electro-optic modulator. 
(j) Pulse stretcher producing group delay dispersion $\Phi_2$ for $\mathscr{F}$. 
(h) A femtosecond laser (\SI{100}{\femto\second}) beam is split with a small power fraction optically driving time lenses, while the rest undergoes spectral filtering (4f) to produce the signal pulse (ca. \SI{10}{\pico\second}). 
(k) Quadruple-pass delay line. 
(N)PBS \textendash{} (non) polarizing beamsplitter,
HWP (QWP) \textendash{} half-(quarter-)waveplate,
amp \textendash{} amplifier,
piezo \textendash{} piezoelectric actuator,
BPF \textendash{} bandpass filter,
EOM \textendash{} electro-optical modulator,
pol ctrl \textendash{} fiber polarization controller,
PID \textendash{} proportional-integral controller,
SPD \textendash{} single-photon detector,
TDC \textendash{} time-digital converter,
HST \textendash{} histogramming module,
LPF \textendash{} low-pass filter,
MMV \textendash{} monostable multivibrator,
ND \textendash{} neutral-density filter.
}
\label{fig:setup}
\end{figure*}

The experimental implementation of \ulfasr{} is depicted in Fig.~\ref{fig:setup}. The central part of the setup consists of a Mach-Zehnder-type interferometer with electro-optic (inverse) Fourier transformers in each arm, described in more detail in ref.~\cite{lipka2023ultrafast}. Each (inverse) Fourier transformer is a sequence of a pulse stretcher with the group delay dispersion (GDD) of $\Phi_2$, a time lens with a chirp rate of $K=1/\Phi_2$, and another pulse stretcher with a GDD of $\Phi_2$. For the direct Fourier transform we have $K>0$, while for the inverse $K<0$. The frequency of the shortest optical path in the pulse stretchers determines the central frequency of the Fourier transformer.

The interferometer is actively phase-stabilized with an intermittent continuous-wave (CW) reference beam. 
The CW beam and the signal pulses are combined with a $\pi/2$ phase difference by employing the polarization degree of freedom. This way, the interferometer can be stabilized for an equal number of photo counts in each arm with CW, while having a dark port with the signal light.  

The input state is prepared by spectrally filtering \SI{100}{\femto\second} pulses from a Ti:Sapphire laser (SpectraPhysics MaiTai) using a custom pulse shaper. A diffraction grating is followed by a folded unit-magnification telescope with a regulated slit and a retro-reflecting mirror placed in the far field of the grating. Motorized lateral movement of the slit alters $\varepsilon$. 

To obtain an incoherent mix of two spectrally separated pulses, partial measurements are collected with a single pulse shifted by $\pm\varepsilon/2$ and the raw photo counts for opposite signs are combined in post-processing. Pulses with a spectrum of \SI{85}{\giga\hertz} full width at half maximum are used.

\section{Discussion\label{sec:discussion}}

Quantum-inspired super-resolution methods employ the information contained in the phase of complex electric field, lost in the direct imaging with quadratic detectors. Proposed and demonstrated implementations often rely on mode-specific detection in a problem-dependent basis of modes. \ulfasr{} as a method based on image inversion interferometry, fundamentally relies on the demultiplexing of a symmetric and antisymmetric part of the input state. However, \ulfasr{} can be extended to the mode sorting of multiple Hermite-Gauss (HG) modes. The details are discussed in appendix sec.~\ref{sec:mode_sorting} where we present two different schemes. One involves a network of SUSI-like interferometers, similar to what has been recently proposed and theoretically analyzed for temporal modes in ref. \cite{horoshko2023interferometric}. The other is fully parallel. We note that a spectral mode-sorting platform constitutes a fundamental device for among others optical communications and spectro-temporal quantum information processing \cite{Brecht2015}.

An alternative approach to mode sorting and super-resolution based on spectro-temporal shaping has been proposed by Shah \textit{et al.} \cite{Shah2021}. It incorporates a cascade of electro-optic modulators with case-specific arbitrary temporal phase modulation intertwined with spectral dispersers. In comparison with \ulfasr{} the serial approach does not require a stabilized interferometer; however, the requirement of high-bandwidth arbitrary electro-optic modulation is very stringent and experimentally infeasible for tens-of-GHz pulses. In the task of mode sorting, with the increasing number of modes the cascade approach requires a growing number of elements in series, whereas \ulfasr{} would involve an interferometer network with a logarithmically growing depth. Depending on the implementation details these approaches could vastly differ in terms of the total efficiency. 

We note that as a general alternative to pure spectral-temporal domain protocols, spectro-spatial mapping allows spatial techniques to be employed in the spectral domain, including measurement \cite{Lipka:21} and mode sorting \cite{Jastrzebski2024spectrum}.

\begin{figure}[ht!]
\centering\includegraphics[width=0.6\columnwidth]{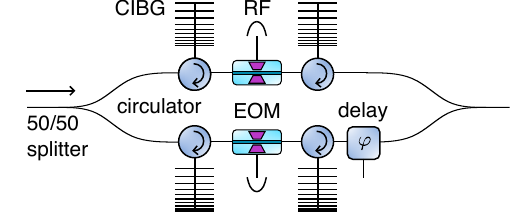}
\caption{
Schematic of a potential \ulfasr{} implementation on a photonic integrated circuit. 
CIBG \textendash{} chirped Bragg grating,
EOM \textendash{} electro-optic modulator,
RF \textendash{} radio-frequency driving signal. Note that the chirp values of CIBGs in the top and bottom arm are opposite.
}
\label{fig:onchip}
\end{figure}

The efficiency of the \ulfasr{} setup is on the order of $0.5\%$, mainly limited by two diffraction-grating-based pulse stretchers in series (total of $1.5\%$ efficiency) implementing part of the Fourier transform. 
Practical applications of \ulfasr{} and extensions to multi-mode sorting would highly benefit from efficient implementations on a photonic chip. We note that all necessary components have been demonstrated including low-loss chirped integrated Bragg gratings (CIBG) \cite{tan2008}, controlled delays \cite{wang2017continuously}, and EOMs \cite{wang2024ultra}. A schematic of an on-chip implementation has been depicted in Fig.~\ref{fig:onchip}

\section{Conclusions\label{sec:conclusions}}
We have experimentally demonstrated \ulfasr{} \textendash{} a spectral-domain super-resolution method based on the principle of image inversion interferometry and compatible with ultrafast pulses with tens-of-GHz bandwidth. The method is aimed at weak ($\ll 1$ photon per coherence time) incoherent light encountered in practical scenarios such as fluorescence spectroscopy or stellar observations. 

Building on the ideas of image inversion interferometry in the position-momentum degree of freedom, our spectral implementation splits the inversion task into two parts. A Fourier transform is placed in one arm of the interferometer and an inverse Fourier transform in the other, ensuring a simple setup and inherent efficiency balance between the arms. The demonstrated experimental implementation requires simple components \textendash{} passive spectral dispersers, a quadratic electro-optic modulation, linear optics, and an active interferometer stabilization. 

The demonstrated \ulfasr{} setup outperforms an equivalent DI measurement with an over two-fold reduction in the required resources for the same estimator variance. 
\ulfasr{} comprises a simple alternative to fairly few other demonstrated quantum-inspired super-resolution methods in the spectral domain. In comparison, no non-linear optics or complex cold-atomic setups are required. Furthermore, the design is compatible with on-chip components and fiber optics, promising practical applications in super-resolution spectroscopy and spectral mode sorting.

\begin{backmatter}
\bmsection{Funding}
Fundacja na rzecz Nauki Polskiej (MAB/2018/4 \textquotedblleft Quantum Optical Technologies\textquotedblright{}); European Regional Development Fund; Narodowe Centrum Nauki (2021/41/N/ST2/02926);

\bmsection{Acknowledgments}
The \textquotedblleft Quantum Optical Technologies\textquotedblright{} project was
carried out within the International Research Agendas programme of the
Foundation for Polish Science co-financed by the European Union under the
European Regional Development Fund.
This research was funded in whole or in part by National Science Centre, Poland 2021/41/N/ST2/02926. 
ML was supported by the Foundation for Polish Science (FNP) via the START scholarship.
We would like to thank K. Banaszek, W. Wasilewski, and M. Mazelanik for the support and discussions.

\bmsection{Disclosures}
The authors declare no conflicts of interest.

\bmsection{Data availability} Data underlying the results presented in this paper are available in ref.~\cite{Data24} (Harvard Dataverse).

\end{backmatter}

\appendix

\section{Mode sorting\label{sec:mode_sorting}}
\begin{figure}[ht!]
\centering\includegraphics[width=1\textwidth]{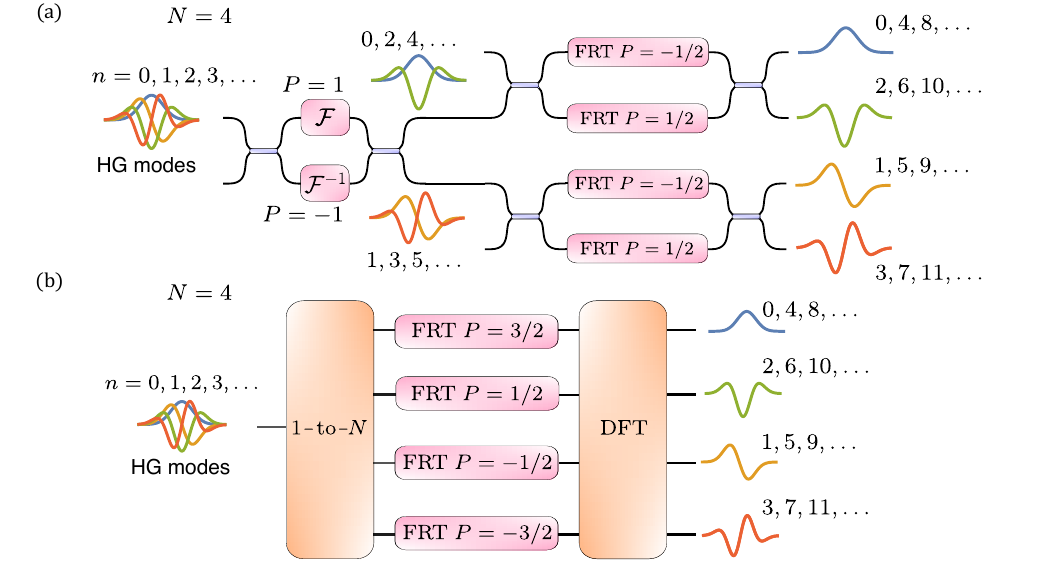}
\caption{
\ulfasr{}-based mode sorting of Hermite-Gauss spectral modes $n$ into $N$ outputs. The $j=0,1,2,\ldots$ output contains $n=j,j+N,j+2N,\ldots$ modes. (a) Architecture based on a network of SUSI-like interferometers with fractional Fourier transforms (FRT) of decreasing orders $P$. (b) Fully parallel approach employing a 1-to-$N$ splitter, FRTs with orders corresponding to equally distributed FRT angles, and a discrete Fourier transform (DFT). 
}
\label{fig:modesorting}
\end{figure}
\ulfasr{}-based mode sorting would use fractional Fourier transforms (FRT) whose implementation is discussed in ref. \cite{lipka2023ultrafast}, and of which Fourier transforms employed in \ulfasr{} are a special case. There are at least two different architectures of mode sorting yet both rely on the fact that an $n$-th HG mode is an eigenfunction of FRT (order $P$) with an eigenvalue $\exp(inP\pi/2)$ which facilitates interferometric separation of HG modes. The two schemes have been depicted in Fig.~\ref{fig:modesorting}.

One solution would employ a network of \ulfasr{}-like interferometers with decreasing FRT orders $P=1/k$ for $k$-th level of the network. Sorting of $N$ modes would require $O(\log N)$ levels and a total of $O(N\log N)$ FRT operations. Decreasing FRT orders alleviate the technical requirements of the FRT implementation, while the losses scale as $O(1/\log N)$. A similar network has been analyzed in ref. \cite{horoshko2023interferometric}.

The other solution would use a 1-to-$N$ beamsplitter followed by $N$ parallel FRTs and a discrete Fourier transform. The FRT angles $\alpha=P\pi/2$ would be equally distributed on a $2\pi$ circle while avoiding the divergent case of $P=2$. In principle, the efficiency of this setup is constant with respect to the number of modes $N$, owing to the parallel configuration of FRTs. 

\ulfasr{} can be regarded as a special case $N=2$ of either of these methods. Both approaches sort the mode numbers $n\bmod N$, hence for $N=2$, \ulfasr{} performs HG sorting into odd $n=0,2,4,\ldots$ and even modes $n=1,3,5,\ldots$.

\section{\ulfasr{} simulation\label{sec:app_simulation}}
An idealized \ulfasr{} setup was simulated with a non-perfect visibility $\mathcal{V}=93\%$ as the only imperfection. The input state was an incoherent mixture of two pulses with a spectral intensity as measured in the experiment and a flat spectral phase. 
The simulation proceeds as follows:
\begin{enumerate}
    \item The experimental spectrum has its background removed by fitting a mixture of a Gaussian and linear functions, and subtracting the latter's best fit.
    \item We take a square root of the spectrum to get an equivalent envelope of the pulse's electric field with a flat spectral phase. Rare small negative values are replaced with zeros.
    \item Spectrum is shifted by $\varepsilon/2$.
    \item Such a signal is then padded with zeros. In parallel a fast Fourier transform (FFT) and its inverse (iFFT) are computed.  
    \item A beamsplitter transformation is performed between FFT ($a$) and iFFT ($b$) arms, accounting for limited visibility 
    \begin{equation}
        \tilde{c}_{\pm}(\varepsilon) = \frac{1}{2}\left[\mathcal{V}\abs{a\pm b}^2 + \left(1-\mathcal{V}\right)\left(\abs{a}^2+\abs{b}^2\right) \right].
    \end{equation}
    \item The results is symmetrized with respect to the change of the $\pm\varepsilon/2$ shift sign 
    \begin{equation}
        c_\pm(\varepsilon)=
        \frac{1}{2}(\tilde{c}_\pm(\varepsilon) +
        \tilde{c}_\pm(-\varepsilon)),
    \end{equation}and the dark port fraction is obtained
    \begin{equation}
        p_\mathrm{sim}(-|\varepsilon)=\frac{c_-(\varepsilon)}{c_+(\varepsilon) +c_-(\varepsilon)}.
    \end{equation}
\end{enumerate}
At this point, Fisher information can be calculated in the same way as from experimental data by first fitting a quadratic function to  $p_\mathrm{sim}(-|\varepsilon)$.

\section{Direct imaging\label{sec:app_di_sim}}
For Gaussian pulses, Fisher information for direct imaging $\mathcal{F}_\mathrm{DI}$ can be evaluated analytically and reads $\varepsilon^2/8$. However, a Gaussian does not describe well our experimental spectrum. We note that minute changes to the spectrum shape can affect $\mathcal{F}_\mathrm{DI}$ drastically. This stems from an unlimited temporal aperture (or equivalently spectral resolution) assumed in our analysis for DI. 

\begin{figure}[ht!]
\centering\includegraphics[width=0.6\columnwidth]{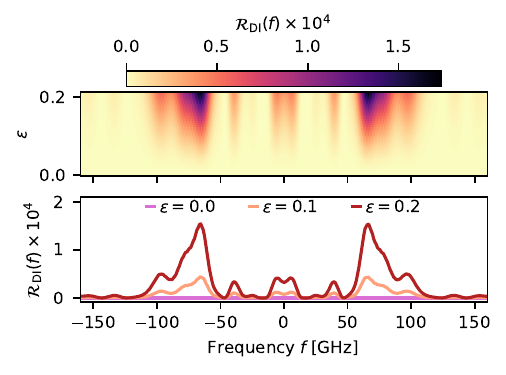}
\caption{
Numerically calculated Fisher information spectral density for $\varepsilon$ estimation with direct imaging.
}
\label{fig:fdidens}
\end{figure}

For a fair comparison, we use the experimentally measured spectrum and calculate the Fisher information numerically. The procedure is as follows:
\begin{itemize}
    \item The spectrum has its background removed by fitting a mixture of a Gaussian and linear functions, and subtracting the latter's best fit.
    \item Two copies of the spectrum are added together, each shifted by $\pm\varepsilon/2$ with opposite signs.
    \item Such a signal is normalized to a unit integral. Let us denote it by $\rho(f|\varepsilon)$.
\end{itemize}
Photo counts density for each frequency $f$ will follow a Poisson distribution with a mean $N\rho(f|\varepsilon)$, where $N$ is the total number of photons and $\int \dd{f} \rho(f|\varepsilon)=1$. 
Fisher information per photon for estimating $\varepsilon$ will be given by
\begin{equation}
    \mathcal{F}_\mathrm{DI} = \int\dd{f}\mathcal{R}_\mathrm{DI}(f),
\end{equation}
where the Fisher information density reads
\begin{equation}
    \mathcal{R}_\mathrm{DI}(f) = \frac
    {\left(\partial_\varepsilon \rho(f|\varepsilon)\right)^2}
    {\rho(f|\varepsilon)}.
\end{equation}
We evaluate this expression numerically and integrate $\mathcal{R}_\mathrm{DI}$ to $\mathcal{F}_\mathrm{DI}$.
The Fisher information density has been depicted in Fig.~\ref{fig:fdidens}.
A quadratic fit gives $\mathcal{F}_\mathrm{DI}=0.24\times\varepsilon^2$.

\section{Setup details and calibration}

\subsection{Pulse preparation}
Signal preparation begins with spectrally filtering \SI{100}{\femto\second} pulses at \SI{798}{\nano\meter} from a Ti:Sapphire laser (Spectra-Physics Mai Tai) with an average power of ca. \SI{4}{\watt} and an \SI{80}{\mega\hertz} repetition rate. The filtering setup, depicted in Fig.~\ref{fig:setup} (g), comprises a diffraction grating (\SI{1800}{lines\per\mm}) far-field imaged onto a rectangular slit with a mirror placed just after the slit, in its near field. The slit has a manually adjustable width and is mounted on a motorized stage controlling the lateral position. The slit width controls the signal bandwidth and the stage movement changes the center frequency of the pulse.
The reflected beam returns to the grating at a slightly different height and leaves through a D-shaped mirror.

For the main experiment, the pulse is set to ca. \SI{85}{\giga\hertz} full width at half maximum.

\subsection{Interferometer stabilization}
During the stabilization period [cf. Fig.~\ref{fig:setup} (c)] time lenses are switched off with an acousto-optical modulator (AOM) deflecting off the optical driving signal (K drv). In turn, a continuous-wave (CW) laser beam is switched on (via an AOM) and probes the phase difference between the interferometer arms. 

As depicted in Fig.~\ref{fig:setup} (a), (f), (j) the CW beam enters the main interferometer together with the signal light. The CW laser (Toptica DL pro 780) matches the central frequency of the pulse stretchers and signal pulses ($\varepsilon=0$). 
The CW beam is combined with the signal on a non-polarizing beamsplitter. 
Circular polarization of the CW and a diagonal polarization of the signal ensures a \SI{90}{\degree} phase delay between these fields. This way, neglecting non-ideal visibility, when the CW light exits through both output ports with equal intensity, the signal will leave only through a single one (the bright port). Hence, we can stabilize the CW intensity difference at the output ports with an unambiguous error signal and high sensitivity. Equivalently, the CW probe is kept at the highest slope of the error signal versus phase, while the signal stays at its extremum.

The CW laser is highly attenuated so that the output ports can be observed with the same superconducting single-photon detectors (ID Quantique ID281) as used for the measurement period. As depicted in Fig.~\ref{fig:setup} (b), photo counts are timestamped by a time-to-digital converter (ID Quantique ID900) which includes a triggered generator (monostable multivibrator) outputting a single pulse of constant width (\SI{50}{\nano\second}) for each registered photo count. These pulses are then processed in an FPGA-based (RedPitaya + custom system) proportional-integral controller (PID). 

Pulses are digitized and low-pass filtered. Their difference constitutes the error signal. Through an amplifier, the output of the PID drives a mirror mounted on a piezo-electric actuator [c.f. Fig.~\ref{fig:setup}(f)].

Fine-tuning of the PID setpoint can correct a small efficiency imbalance in the interferometer arms, slightly mismatched polarization of the CW and signal beams, and imperfect splitting of the NPBS.

During the measurement period, the PID is stopped and holds the last output value. During normal operation, the integral part will sometimes saturate and automatically reset. Such a behavior is required since the stabilized quantity \textendash{} the optical phase difference is periodic. 

Since only AOMs have to switch (which takes at most a few \SI{}{\micro\second}) between the measurement and stabilization periods, we can use short measurement windows (\SI{200}{\micro\second}). Which is a requirement with a large, quickly fluctuating interferometer. 

The experimental sequence is controlled by a sequencer implemented in the FPGA and programmed from the PC.

\subsection{Fourier transformer calibration}
An ideal Fourier transformer requires a matched GDD of both pulse stretchers. We cross-calibrated 4 pulse stretchers \textendash{} 2 positive-dispersion for direct Fourier transform and 2 negative-dispersion for the inverse. Each positive-dispersion stretcher was put in series with each negative-dispersion stretcher in a single arm of a Mach-Zehnder-type interferometer. The other arm consisted of a regulated delay. A test pulse with a spectral width of a few hundred GHz (maximal possible with our 4f pulse shaper) was sent into both arms. With a spectrometer, we observed fringes and adjusted the delay to cancel a linear component in the spectral phase difference between the arms. One of the stretchers $\Phi_2$ was then adjusted by moving a platform with the grating, depicted in Fig.~\ref{fig:setup} (i).

The second requirement $K=1/\Phi_2$ was accomplished by adjusting the time lens chirp rate $K$ via a programmable attenuator controlling the amplitude of the EOM driving signal. The time lens details are depicted in Fig.~\ref{fig:setup} (h). A pair of temporally separated coherent pulses (produced with an auxiliary interferometer with a delay line) was sent through the Fourier transformer under calibration. The output was observed with a spectrometer. With $K=0$ spectral fringes can be observed. As $K$ is increased the pulses spectrally separate. $K$ was set such that the observed separation matched a theoretically calculated one for $K=1/\Phi_2$.

In turn, $\Phi_2$ measurements were carried with the stationary phase point method \cite{sainz1994real}. Its implementation is described in detail in ref.~\cite{lipka2023ultrafast}.

\subsection{Single-photon histograms}
The TDC implements a histogramming module which is triggered from the FPGA sequencer. Histograms are set to have \SI{1}{\micro\second} wide bins and around \SI{3.5}{\milli\second} are collected in each shot. The sequence is repeated at ca. \SI{100}{\hertz}. PC controls the collection and storage of histograms, settings of the FPGA sequencer, and position of the stepper motor corresponding to $\varepsilon$. The latter via a custom ARM-based controller (stm32f103c8t6).

\bibliography{ulfasr}

\end{document}